\documentclass[aps,showpacs,twocolumn]{revtex4}
\usepackage{amsfonts}
\usepackage{amsmath}
\usepackage{amssymb}
\usepackage{graphicx}
\usepackage{bm}
\input{tcilatex}

\begin{document}

\title{Hydrodynamic boundary effects on thermophoresis of confined colloids}
\author{Alois W\"{u}rger}
\affiliation{Laboratoire Ondes et Mati\`{e}re d'Aquitaine, Universit\'{e} de Bordeaux \&
CNRS, 33405 Talence, France}

\begin{abstract}
We study hydrodynamic slowing-down of a particle moving in a temperature
gradient perpendicular to a wall. At distances much smaller than the
particle radius, $h\ll a$, lubrication approximation leads to the reduced
velocity $u/u_{0}=3\frac{h}{a}\ln \frac{a}{h}$, with respect to the bulk
value $u_{0}$. With Brenner's result for confined diffusion, we find that
the trapping efficiency, or effective Soret coefficient, increases
logarithmically as the particle gets very close to the wall. This provides a
quantitative explanation for the recently observed enhancement of
thermophoretic trapping at short distances. Our discussion of parallel and
perpendicular thermophoresis in a capillary, reveals a very a good agreement
with five recent experiments on charged polystyrene particles.
\end{abstract}

\maketitle

The motion of a colloid close to a solid boundary is strongly influenced by
hydrodynamic interactions. Thus the like-charge attractions observed for
confined colloidal assemblies \cite{Lar97}, were shown to arise from
hydrodynamic fluctuations \cite{Squ00}. Similarly, a surface-active particle
with a flow field perpendicular to the wall, induces lateral advection of
nearby neighbors and cluster formation \cite{Tra96,Yeh97,Wei08,Dil09,Mor10}.
More recently, the collision patterns observed for self-propelling Janus
particles close to a wall \cite{Vol11}, were related to hydrodynamic
interactions. Quite generally, the latter are relevant where surface forces
and\ confined geometries are combined for sieving \cite{Cuc14}, trapping 
\cite{Bra14,Che15}, and assembling colloidal beads \cite{The12}.

A generic example is provided by a surface-active particle moving towards a
wall. If hydrodynamic effects on Brownian motion are well understood in
terms of Brenner's solution for confined diffusion \cite{Bre61}, this is not
the case for the drift velocity $u$. At distances $h$\ much larger than the
particle radius $a$, electrophoresis slows down by the factor $u/u_{0}=1-%
\frac{5}{8}a^{3}/h^{3}$ \cite{Keh85}. At short distances $h<a$, the
wall-solvent-particle permittivity contrast strongly alters the local
electric field; this electric coupling is difficult to separate from
hydrodynamic interactions; a similar effect influences diffusiophoresis \ \ 
\cite{Cha08}. A more favorable situation occurs for thermophoresis, where
the drift velocity is proportional to the temperature gradient \cite%
{Pia08,Wue10}: Since the heat conductivity of silica or polystyrene (PS)
particles is not very different from that of water, the thermal gradient is
hardly affected by the presence of the wall \cite{Mor10}, and velocity
changes can be unambiguously attributed to hydrodynamic interactions.

Here we study the vertical motion of a particle that is confined to the
upper half-space $z\geq 0$, as illustrated in Fig. 1; applications are
thermophoresis across a capillary and self-propelling Janus particles that
preferentially orient toward the wall, or \textquotedblleft
pullers\textquotedblright\ \cite{Sch15}. In the steady state, drift and
diffusion currents cancel each other, $-uc-D\nabla c=0$, and the particle
concentration satisfies%
\begin{equation}
-\nabla \ln c=\frac{u}{D}.  \label{4}
\end{equation}%
At large distances $h\gg a$, there are no boundary effects and Eq. (\ref{4})
is readily integrated, $c=c_{0}e^{-h/\ell _{0}}$, with the trapping length $%
\ell _{0}=D_{0}/u_{0}$ \cite{Esl14}. As the particle approaches the wall,
both drift and diffusion are slowed down by hydrodynamic coupling.\ As a
main result, we calculate the velocity reduction factor $u/u_{0}$ in
lubrication approximation ($h<a$) and by the method of reflections ($h>a$);
with Brenner's expression for the diffusion coefficient \cite{Bre61}, we
evaluate (\ref{4}) for both limiting cases.

The present work was partly motivated by the recent observation that
thermophoretic trapping at very short distances is much stronger than in the
bulk \cite{Hel15}. When relating (\ref{4}) to the Soret coefficient $S_{T}$
through $u/D=S_{T}\nabla T$, our results agree quantitatively with the data
and provide strong evidence that the observed increase of $S_{T}$ is of
hydrodynamic origin.

\begin{figure}[ptb]
\includegraphics[width=\columnwidth]{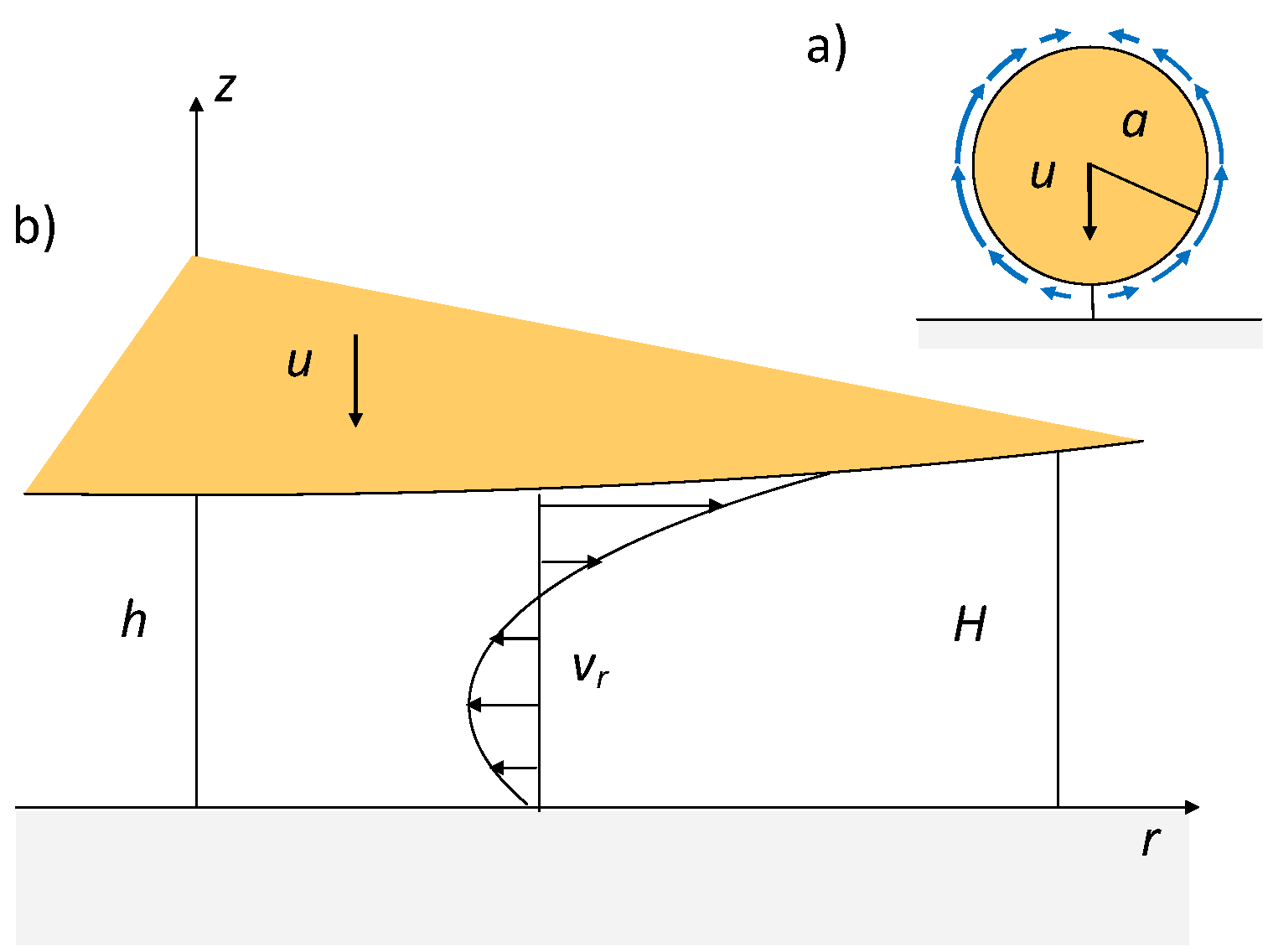}
\caption{Schematic view of a particle
moving towards a confining wall at velocity $u$. a) The arrows along the
particle surface indicate the slip velocity $v_{s}$ induced by thermodynamic
forces. b) In the narrow slit of width $H(r)=h+a-\protect\sqrt{a^{2}-r^{2}}$%
, the vertical particle motion and the outward slip velocity result in an
intricate radial flow profile $v_{r}$.}
\end{figure}

\textit{Hydrodynamic boundary effects. }Particle drift arises from the
effective slip velocity $v_{s}$ induced by thermal non-equlibrium properties
of the boundary layer \cite{Wue10,And89}; with the notation of Fig. 1 one
has $v_{s}=\frac{3}{2}u_{0}(r/a)$. As the particle approaches the boundary,
the wall squeezes the flow field and thus reduces the drift velocity to the
value $u$.

At very short distances, the flow in the slit is well described by
lubrication approximation, with the radial velocity\ field\ 
\begin{equation}
v_{r}(z)=v_{s}\left( \frac{z}{H}-\frac{3z(H-z)}{H^{2}}\right) +3u\frac{r}{H}%
\frac{z(H-z)}{H^{2}}.  \label{16}
\end{equation}%
The first term arises from the slip velocity $v_{s}$, and satisfies the
conditions $v_{r}|_{0}=0$ at the solid boundary and $v_{r}|_{H}=v_{s}$ at
the particle surface $H=h+a-\sqrt{a^{2}-r^{2}}$. Integrating over $z$, one
finds that its net flow vanishes. The second term accounts for the particle
velocity $u$; one readily verifies that the vertical volume flow $\pi r^{2}u$
within a radius $r$, is cancelled by the radial flow through a cylinder of
radius $r$ and height $H$, that is, the $z$-integral of $2\pi rv_{r}$.

The relation between the particle velocity $u$ and the surface property $%
v_{s}$, is established by noting that there is no force $F$ between\
particle and wall. From the radial component of Stokes' equation, $\partial
_{r}P=\eta \partial _{z}^{2}v_{r}$, we obtain the pressure gradient $%
\partial _{r}P=6\eta v_{s}/H^{2}-6\eta ur/H^{3}$ which, upon integration,
gives $P(r)$. Performing its surface integral along the wall and using that
the diagonal component of the viscous stress vanishes, $\sigma _{zz}=0$, we
calculate \cite{SM} 
\begin{equation}
F=\int dSP(r,z=0)=0.  \label{17}
\end{equation}%
The second equality expresses the fact that there is no mechanical or
`thermophoretic' force acting on the particle.

\begin{figure}[ptb]
\includegraphics[width=\columnwidth]{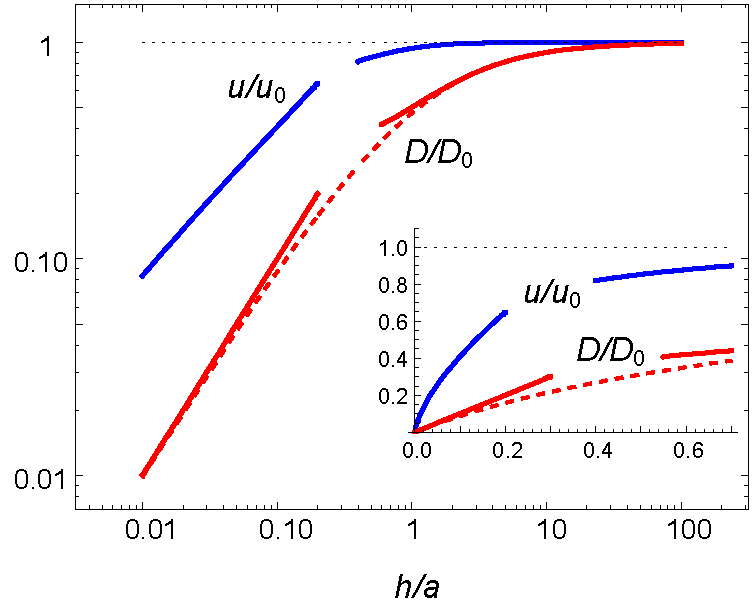}
\caption{Reduced drift velocity $u/u_{0}$\ and diffusion
coefficient $D/D_{0}$ for a particle moving toward a wall, as a function of
the relative distance $\hat{h}=h/R$. The curves at small $\hat{h}$ are given
by (\protect\ref{18}) and (\protect\ref{20}), those at large $\hat{h}$ by (%
\protect\ref{21}) and (\protect\ref{22}). The dashed line is calculated from
the first hundred terms of Brenner's exact series for $D/D_{0}$ \protect\cite%
{Bre61}. At all distances, the diffusion coefficient is more strongly
reduced than the drift velocity. The inset shows the same at linear scale,
thus highlighting the linear law $D\propto \hat{h}$ and the logarithmic
corrections for $u$ at short distances}
\end{figure}

The condition (\ref{17}) provides a relation between $u$ and $u_{0}$, and
thus quantifies the hydrodynamic effects on the drift velocity,%
\begin{equation}
\frac{u}{u_{0}}=\frac{h}{a}\phi (h/a)\ \ \ \ \ (h\ll a),  \label{18}
\end{equation}%
with 
\begin{equation}
\phi (\hat{h})=3(1+\hat{h})\frac{(2+6\hat{h}+3\hat{h}^{2})\ln \frac{\hat{h}+1%
}{\hat{h}}-\frac{3}{2}(3+2\hat{h})}{2+9\hat{h}+6\hat{h}^{2}-6\hat{h}(1+\hat{h%
})^{2}\ln \frac{\hat{h}+1}{\hat{h}}}  \label{19}
\end{equation}%
and the shorthand notation $\hat{h}=h/a$. For very small distances, $\hat{h}%
\leq \frac{1}{100}$, this expression simplifies to $\phi =-3(\ln \hat{h}%
+9/4) $. With (\ref{18}) we have an explicit expression for the velocity
profile (\ref{16}).

Now we turn to the case where the distance exceeds the particle size, $h>a$.
Following Keh and Anderson \cite{Keh85}, we start from the velocity field in
a bulk liquid and evaluate the first reflection at the wall \cite{SM}. The
resulting correction to the particle velocity vanishes as $h^{-3}$, 
\begin{equation}
\frac{u}{u_{0}}=1-\frac{1}{2}\frac{a^{3}}{(h+a)^{3}}\ \ \ (h>a).  \label{20}
\end{equation}%
A slightly larger correction, with a prefactor $\frac{5}{8}$ instead of $%
\frac{1}{2}$, was found for the electrophoretic mobility \cite{Keh85}.\ The
difference of $\frac{1}{8}$ arises from the deformation of the electric
field by the low-permittivity particle and by the conducting wall. In the
case of thermophoresis, the corresponding effect on the local temperature
gradient is small, because of the relatively weak thermal conductivity
contrast at the particle-solvent-wall interfaces \cite{SM}. A more complex
situation occurs if ion currents are relevant for the slip velocity, e.g.,
through the Seebeck effect \cite{Esl14,Wue08} or a permittivity change due
to phase separation \cite{But12,Wue15}.

Fig. 2 shows the reduced velocity $u/u_{0}$ as a function of distance; it
changes rather little for $h>a$, but drops to zero as $h\rightarrow 0$.\ For
comparison we also plot the corresponding expressions $D/D_{0}$ for the
diffusion coefficient. At small distances, the lubrication approximation
results in the well-known linear variation with $h$, 
\begin{equation}
D/D_{0}=h/a\ \ \ \ \ \ (h<a),  \label{21}
\end{equation}%
whereas to third order in the inverse distance, the reflection method
results in 
\begin{equation}
\frac{D}{D_{0}}=1-\frac{9}{8}\frac{a}{h+a}+\frac{1}{2}\frac{a^{3}}{(h+a)^{3}}%
\ \ \ \ \ (h>a).  \label{22}
\end{equation}%
The dashed curve gives Brenner's exact formula \cite{Bre61}. As a general
rule, hydrodynamic slowing down is significantly stronger for diffusion, as
a consequence of the long-range velocity field accompanying Browian motion.%

\begin{figure}[ptb]
\includegraphics[width=\columnwidth]{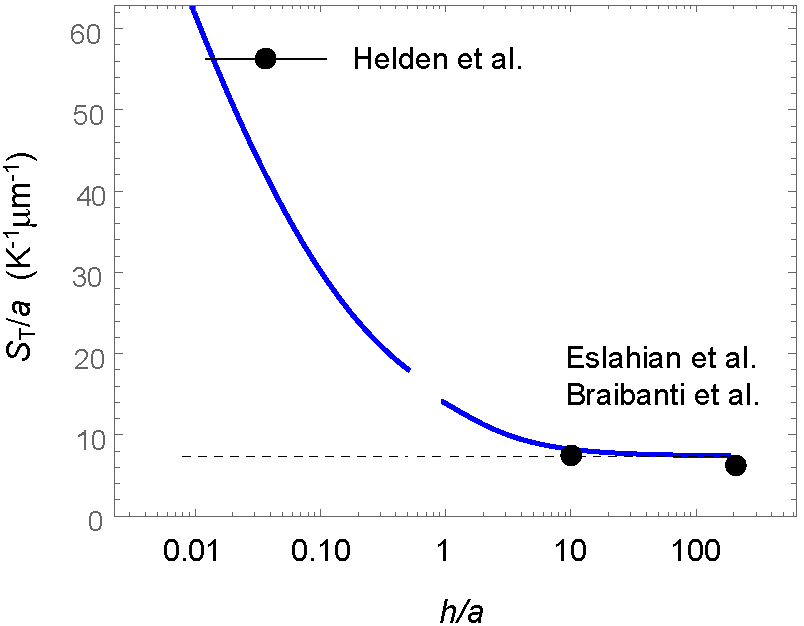}
\caption{Hydrodynamic effects on the
effective Soret coefficient $S_{T}/a$. The solid lines show the results from
lubrication approximation (\protect\ref{24}) and the method of reflection
from (\protect\ref{30}); the dashed line gives the bulk value $%
S_{T}^{0}/a=7.4$ K$^{-1}\protect\mu $m$^{-1}$ \protect\cite{SM}. The
experimental points are from Table I.}
\end{figure}

\textit{Thermophoretic trapping.} In view of a recent Soret experiment by
Helden et al. \cite{Hel15}, we discuss confined thermophoresis, where the
stationary distribution (\ref{4}) defines the Soret coefficient $S_{T}$
through $u/D=S_{T}\nabla T$. Assuming a constant temperature gradient, one
obtains hydrodynamic effects as the ratio of the correction factors for
drift and diffusion. In lubrication approximation, this results in 
\begin{equation}
S_{T}=S_{T}^{0}\phi (\hat{h})\ \ \ \ (\hat{h}<1),  \label{24}
\end{equation}%
whereas in the opposite limit we have 
\begin{equation}
S_{T}=S_{T}^{0}\frac{1-\frac{1}{2}\frac{1}{(1+\hat{h})^{3}}}{1-\frac{9}{8}%
\frac{1}{1+\hat{h}}+\frac{1}{2}\frac{1}{(1+\hat{h})^{3}}}\ \ \ \ (\hat{h}>1).
\label{30}
\end{equation}%
In Fig. 3 we plot the Soret coefficient as a function of the reduced
distance $h/a$. As the particle gets closer to the wall, trapping is
enhanced by hydrodynamic interactions, the Soret coefficient increases with
respect to the bulk value, and at $\hat{h}\rightarrow 0$ diverges
logarithmically.

\begin{figure}[ptb]
\includegraphics[width=\columnwidth]{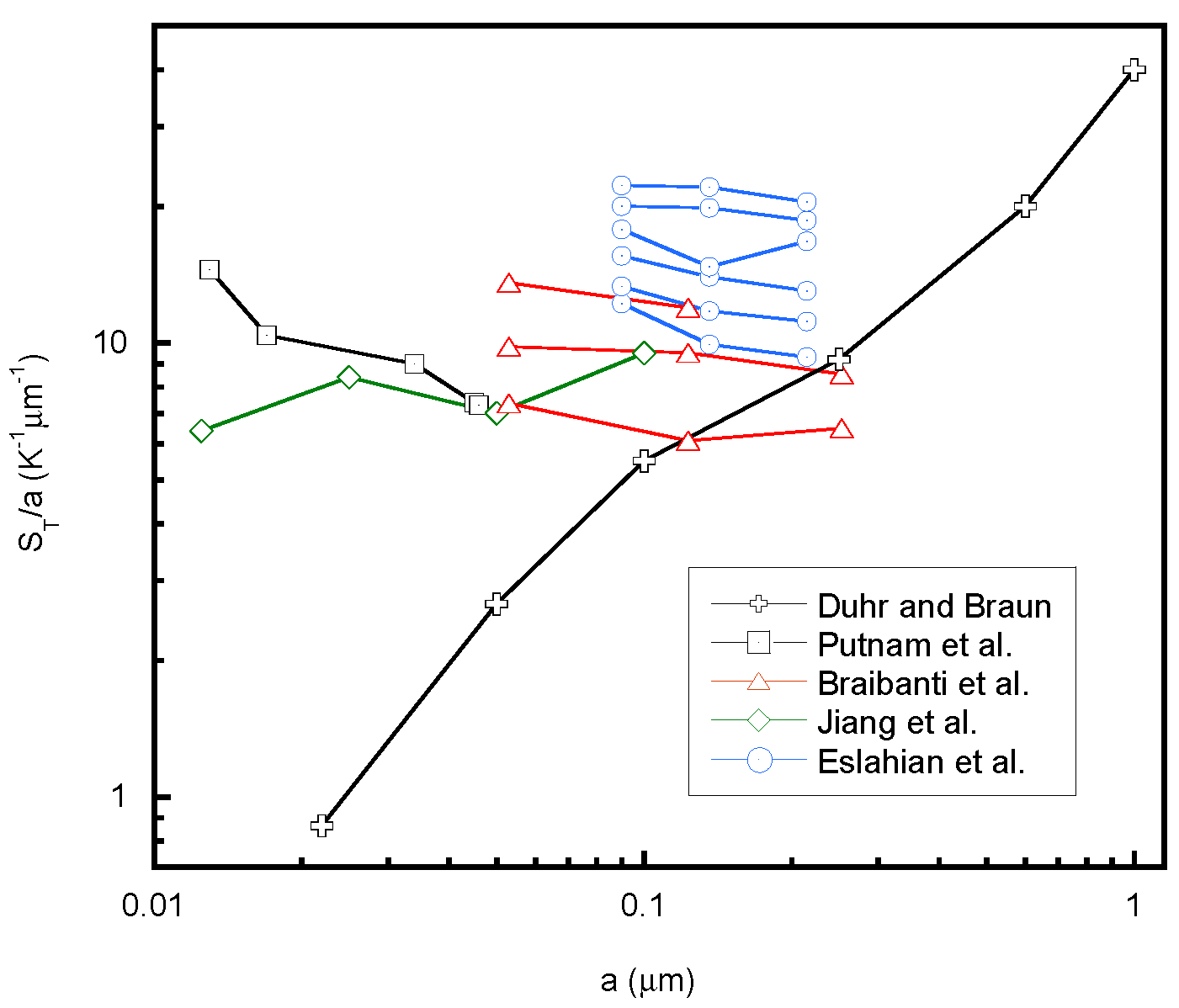}
\caption{Size dependence of the Soret
coefficient $S_{T}$. For the data of five experiments on polystryene
particles, the ratio of $S_{T}$ and the particle radius $a$ is plotted as a
function of $a$. The data (+) are taken at room temperature, Duhr and Braun 
\protect\cite{Duh06}; ($\square $) above 35  $%
{{}^\circ}%
$C, Putnam et al. \protect\cite{Put07}; ($\triangle $) at 25, 35, 45 $%
{{}^\circ}%
$C, Braibanti et al. \protect\cite{Bra08}; ($\diamondsuit $) Jiang et al. 
\protect\cite{Jia09}; ($\bigcirc $) at 28, 31, 35, 39, 44, 47 $%
{{}^\circ}%
$C, Eslahian et al. \protect\cite{Esl14}. The lines connect data at constant
temperature. Details are given in \protect\cite{SM}.}
\end{figure}

\begin{table}[b]
\caption{Soret data for polystyrene particles in capillaries with a
perpendicular temperature gradient \protect\cite{Hel15,Esl14,Bra08}. The
numbers $S_{T}/a$ give the data shown in Fig. 5 \protect\cite{Bra08}, or
their extrapolation to $T=25%
{{}^\circ}%
$ \protect\cite{Esl14}; that of Helden et al. is taken from Fig. 4 of 
\protect\cite{Hel15}. The values for the distance $h$ correspond to the
range where data are taken \protect\cite{Hel15} or, for weak trapping, to $%
\ell _{0}=D_{0}/u_{0}$ \protect\cite{Esl14,Bra08}; the value for Ref. 
\protect\cite{Bra08} indicates a lower bound. For a more detailed analysis,
see \protect\cite{SM}.}%
\begin{tabular}{|l|c|c|c|}
\hline
$T=25%
{{}^\circ}%
$\ C & $%
\begin{array}{c}
S_{T}/a \\ 
(\text{K}^{-1}\mu \text{m}^{-1})%
\end{array}%
$ & $%
\begin{array}{c}
h \\ 
(\mu \text{m})%
\end{array}%
$ & $%
\begin{array}{c}
h/a \\ 
\ 
\end{array}%
$ \\ \hline
Helden et al. \cite{Hel15} & 56 & $0.03-0.3$ & $0.012-0.12$ \\ \hline
Eslahian et al. \cite{Esl14} & 7.4 & $2$ & 9 \\ \hline
Braibanti et al. \cite{Bra08} & 6.5 & $>50$ & \TEXTsymbol{>} 200 \\ \hline
\end{tabular}%
\linebreak \linebreak
\end{table}

The bulk Soret coefficient.varies linearly with the particle radius, $%
S_{T}^{0}\propto a$, due to the inverse variation of the Stokes-Einstein
coefficient $D_{0}=k_{B}T/(6\pi \eta a)$\ \cite{Ein06} and the constant
drift velocity $u_{0}$ \cite{And89}. In order to facilitate the comparison
of Soret data for particles of different radius, we plot the ratio $S_{T}/a$%
. Our findings provide a quantitative explanation for\ the data of Helden et
al. \cite{Hel15}: For polystyrene particles ($a=2.5\mu $m) very close to a
wall ($h<0.3\mu $m), these authors\ reported $S_{T}=140$ K$^{-1}$ at room
temperature; the reduced value $S_{T}/a$ is seven times larger than those
reported in previous experiments on particles at large distances; see Fig. 3
and Table I. The quantitative agreement with the present theory provides
strong evidence that the enhanced trapping is of hydrodynamic origin. This
is corroborated by the similar temperature series observed at small \cite%
{Hel15} and large distances \cite{Esl14,Bra08}.

Since the linear size dependence of $S_{T}^{0}$ is essential for the above
argument, we recall its theoretical foundation and experimental
confirmation. If the Stokes-Einstein coefficient needs no further
discussion, a few words are in order concerning $u_{0}$. As first shown by
von Smoluchowski in his study of thin-boundary layer electrophoresis \cite%
{Smo03}, the equilibrium between surface forces and viscous stress is
independent of the particle radius, and so is the velocity $u_{0}$. Later
on, Derjaguin generalized this argument to motion driven by composition and
temperature gradients \cite{Der87}. The law $u_{0}=$const. ceases to be
valid for no-stick boundary conditions with large Navier slip length and for
particles smaller than the Debye length \cite{Wue10}; yet none of these
cases is relevant for the systems considered here.

Fig. 4 shows Soret data $S_{T}/a$ as a function of $a$, measured for PS
particles in five experiments. In the setup of Refs. \cite{Esl14,Put07,Bra08}
the temperature gradient is perpendicular to the boundary as in Fig. 1; a
parallel configuration is used in \cite{Jia09,Duh06}, with the particles
moving along the capillary. The data of \cite{Esl14,Put07,Jia09,Bra08} show
the behavior $S_{T}/a=\mathrm{const}$. expected for large $h$, and even
their absolute values agree well with each other. A constant ratio was also
observed for surfacted microemulsion droplets \cite{Vig07}. On the contrary,
Duhr and Braun reported a linear variation $S_{T}/a\propto a$ over two
orders of magnitude \cite{Duh06}.

\begin{figure}[ptb]
\includegraphics[width=\columnwidth]{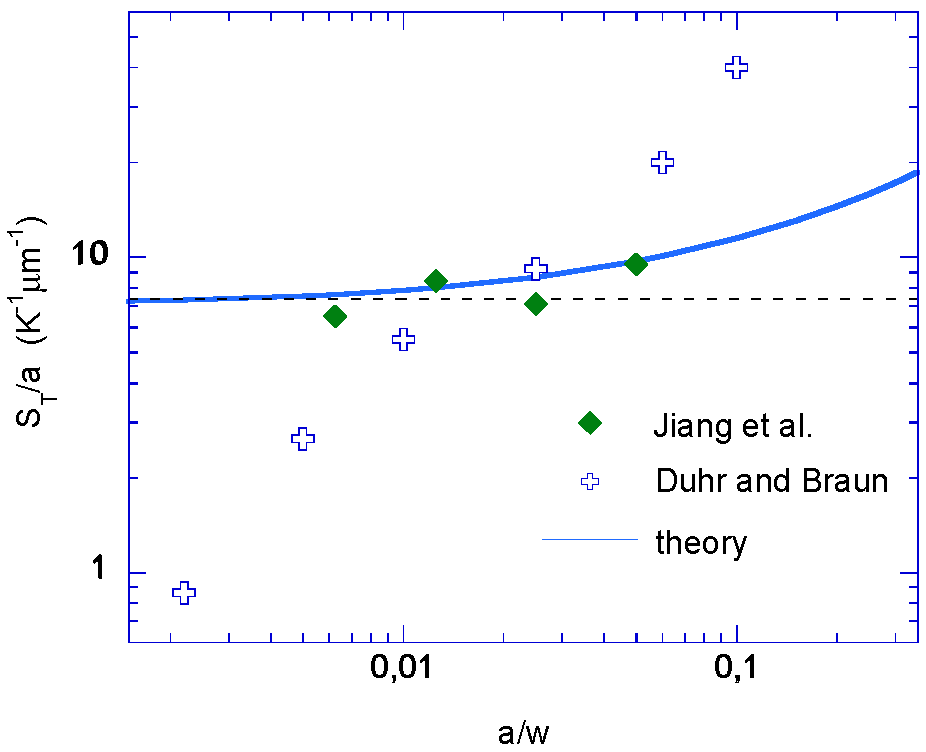}
\caption{
Thermophoretic trapping parallel to the capillary. We show $S_{T}/a$ as a
function of $a/w$. The data of Duhr and Braun are taken in a capillary of
width $w=10\protect\mu $m \protect\cite{Duh06}, and those of Jiang et al. in 
$w=2\protect\mu $m \protect\cite{Jia09}. The solid line is calculated from (%
\protect\ref{26}) where hydrodynamic effects are accounted for by Oseen's
model for the parallel diffusion coefficient. The dashed line indicates the
bulk value $S_{T}^{0}/a=7.4$ K$^{-1}\protect\mu $m$^{-1}.$}
\end{figure}

\textit{Motion parallel to the capillary.} In view of this discrepancy, we
complete our discussion of hydrodynamic effects by considering
thermophoresis along the boundaries, where confinement is expressed in terms
of the ratio of the particle radius $a$ and the width $w$ of the capillary.\
In view of the experiments \cite{Jia09,Duh06} we consider the perturbative
range $a\ll w$ only. In Oseen's model for parallel diffusion, confinement
reduces the Stokes-Einstein coefficient of a particle at vertical position $%
z $ according to $D_{0}/D_{\Vert }=1+\frac{9}{8}\frac{a}{z}+\frac{9}{8}\frac{%
a}{w-z}$ \cite{Bre61}. The thermophoretic velocity is hardly affected by the
walls, in leading order we have $u/u_{0}=1$. Taking the position average in
the interval $[a,w-a]$, we find 
\begin{equation}
\frac{S_{T}}{S_{T}^{0}}=1+\frac{9}{4}\frac{a}{(w-2a)}\ln \frac{w-a}{a}\ \ \
\ (a\ll w).  \label{26}
\end{equation}

In Fig. 5 we plot the reduced Soret data from Refs. \cite{Jia09,Duh06} as a
function of $a/w$, and compare with the theoretical expression (\ref{26}).\
If the four data points of Jiang et al. \cite{Jia09} agree well with theory,
this is not the case for those of Duhr and Braun: The Soret effect of the
biggest particles ($a=1$ $\mu $m) is three times stronger, whereas that of
the smallest one ($a=22$ nm) is by one order of magnitude too weak. This
discrepancy can not be explained by higher-order terms in (\ref{26}) or by
additional effects such as thermoosmosis along the capillary. For a
discussion of experimental issues, see Ref.\ \cite{Bra08}.

\textit{Conclusion.} \ We have studied hydrodynamc effects on the velocity
and the confinement of colloids approaching a wall. At very small distances $%
\hat{h}\ll 1$, lubrication approximation results in the reduction factor $%
u/u_{0}=-3\hat{h}\ln \hat{h}$, which is by a logarithmic factor larger than
that of the diffusion coefficient,\ $D/D_{0}=\hat{h}$.\ 

Our theory provides a quantitative explanation for a recent experiment on
confined thermophoresis \cite{Hel15}, where the measured Soret coefficient
is by almost one order of magnitude larger than expected from previous
studies without confinement. This confirms Derjaguin's approach to thermal
motion which is based on surface forces and hydrodynamics.

More generally, our findings show that hydrodynamic coupling strongly
affects the motion of surface-active colloids in micron-size capillaries or
close to solid boundaries.

Stimulating discussions with L.\ Helden are gratefully acknowledged. This
work was supported by Agence Nationale de la Recherche through contract
ANR-13-IS04-0003,


\begin{thebibliography}{99}
\bibitem{Squ00} T.M. Squires, M.P. Brenner, Phys. Rev. Lett. \textbf{85},
4976 (2000)

\bibitem{Lar97} A.E.\ Larsen, D.\ G.\ Grier, Nature \textbf{385}, 230 (1997)

\bibitem{Tra96} M. Trau, D. Saville, and A. I. Askay, Science \textbf{272},
706 (1996)

\bibitem{Yeh97} S.R.\ Yeh, M.\ Seul, B.I.\ Shraiman, Nature \textbf{386}, 57
(1997)

\bibitem{Wei08} F.M. Weinert, D. Braun, Phys. Rev. Lett. \textbf{101},
168301 (2008)

\bibitem{Dil09} R. Di Leonardo, F. Ianni, G.\ Ruocco, Langmuir \textbf{25},
4247 (2009)

\bibitem{Mor10} J.\ Morthomas, A.\ W\"{u}rger, Phys. Rev.\ E \textbf{81},
051405 (2010)

\bibitem{Vol11} G. Volpe, I. Buttinoni, D. Vogt, H.-J. K\"{u}mmerer, and C.
Bechinger, Soft Matter \textbf{7}, 8810 (2011)

\bibitem{Spa12} S.E. Spagnolie and E. Lauga, J. Fluid Mech. \textbf{700},
105.(2012)

\bibitem{Sch15} K.\ Schaar, A.\ Z\"{o}ttl, H.\ Stark, Phys. Rev.\ Lett. 
\textbf{115}, 038101 (2015)

\bibitem{Cuc14} A. Cuche, A. Canaguier-Durand, E. Devaux, J. A. Hutchison,
C. Genet, T.W. Ebbesen, Nano Lett. \textbf{13}, 4230 (2013)

\bibitem{Che15} J. Chen, Z. Kang, S.K. Kong, and H.-P. Ho, Optics Lett. 
\textbf{40}, 3926 (2015)

\bibitem{Bra14} M. Braun, A. W\"{u}rger and F. Cichos, Phys. Chem. Chem.
Phys. \textbf{16}, 15207 (2014)

\bibitem{The12} I. Theurkauff, C. Cottin-Bizonne, J. Palacci, C. Ybert, L.
Bocquet, Phys. Rev. Lett.\textbf{\ 108}, 268303 (2012)

\bibitem{Bre61} H.\ Brenner, Chem.\ Eng. Sci. \textbf{16}, 242\ (1961)

\bibitem{Keh85} H.J.\ Keh, J.L.\ Anderson, J. Fluid Mech. \textbf{153}, 417\
\ (1985)

\bibitem{Cha08} Y.C. Chang, H.J. Keh, J. Colloid Interf. Sci. \textbf{322},
634 (2008)

\bibitem{Pia08} R.\ Piazza, Soft Matter \textbf{4}, 1740 (2008).

\bibitem{Wue10} A. W\"{u}rger, Rep.\ Prog. Phys. \textbf{73}, 126601 (2010)

\bibitem{Esl14} K.A. Eslahian, A. Majee, M. Maskos, A. W\"{u}rger, Soft
Matter \textbf{10}, 1931 (2014)

\bibitem{Hel15} L.\ Helden, R. Eichhorn, C.\ Bechinger, Soft Matter \textbf{%
11}, 2379\ \ (2015)

\bibitem{SM} See Supplemental material for technical details

\bibitem{Wue08} A. W\"{u}rger, Phys. Rev. Lett. \textbf{101}, 108302 (2008)

\bibitem{But12} I. Buttinoni, G. Volpe, F. K\"{u}mmel, G. Volpe, and C.
Bechinger, J. Phys. Condens. Matter \textbf{24}, 284129 (2012)

\bibitem{Wue15} A. W\"{u}rger, Phys. Rev. Lett. \textbf{115}, 188304 (2015)

\bibitem{Ein06} A. Einstein, Annal. Phys. \textbf{19}, 371 (1906)

\bibitem{And89} J.L. Anderson, Ann. Rev. Fluid Mech. \textbf{21}, 61 (1989)

\bibitem{Smo03} M.\ von Smoluchowski, Bull. Int. Acad. Sci. Cracovie \textbf{%
184}.(1903)

\bibitem{Der87} N.V. Churaev, B.V. Derjaguin, V.M. Muller, \textit{Surface
Forces}, \ Plenum Publishing Corporation (New York 1987)

\bibitem{Put07} S.A. Putnam, D.G.\ Cahill, G.C.L. Wong, Langmuir \textbf{23}%
, 9221 (2007)

\bibitem{Bra08} M. Braibanti, D. Vigolo, R. Piazza, Phys. Rev. Lett.\textbf{%
\ 100}, 108303 (2008).

\bibitem{Jia09} H.-R. Jiang, H.\ Wada, N. Yoshinaga and M. Sano, Phys. Rev.
Lett. \textbf{102}, 208301 (2009)

\bibitem{Duh06} S.\ Duhr, D.\ Braun, Phys. Rev. Lett. \textbf{97}, 168301
(2006)

\bibitem{Vig07} D. Vigolo, S.\ Brambilla, R.\ Piazza, Phys. Rev. E \textbf{75%
}, 040401 (2007)
\end{thebibliography}
\end{document}